\documentclass[11pt]{article}
\usepackage{xspace}
\usepackage{graphicx}
\usepackage{amsmath}
\usepackage{amssymb}
\usepackage{color}

\definecolor{Red}{rgb}{1,0,0}
\definecolor{Green}{rgb}{0,1,0}
\definecolor{Blue}{rgb}{0,0,1}
\definecolor{Black}{rgb}{0,0,0}

\def\beq{\begin{equation}}
\def\eeq#1{\label{#1}\end{equation}}
\def\eeqn{\end{equation}}

\def\beqa{\begin{eqnarray}}
\def\eeqa#1{\label{#1}\end{eqnarray}}
\def\eeqan{\end{eqnarray}}

\let\bar=\overbar

\def\Dslash{\not{\hbox{\kern-4pt $D$}}}
\def\dslash{\not{\hbox{\kern-2pt $\del$}}}

\def\msb{{\bar{\ssstyle M \kern -1pt S}}}

\def\Title#1{\begin{center} {\Large {\bf #1} } \end{center}}

\begin{document}

\Title{Why a NESSiE-like experiment at SBL is needed?}

\bigskip\bigskip


\begin{raggedright}  

{\it Laura Pasqualini, for the NESSiE Collaboration \\
University of Bologna and INFN\\
Bologna, ITALY}\\

\end{raggedright}
\vspace{1.cm}

{\small
\begin{flushleft}
\emph{To appear in the proceedings of the Prospects in Neutrino Physics Conference, 15 -- 17 December, 2014, held at Queen Mary University of London, UK.}
\end{flushleft}
}

\section{Introduction}

Nowadays there are several experimental results that may hint to a deviation from the coherent picture of the standard three neutrino oscillation model. These results, not statistically significant on a single basis, could be explained with the hypothesis of the existence of additional neutrino mass eigenstates at eV mass scale, namely sterile neutrinos \cite{sterile}. In particular there are strong tensions between $\nu_{e}$ disappearance and $\nu_{e}$ appearance results in experiment at Short-Baseline (SBL). The main source of tension arises from the absence of any $\nu_{\mu}$ disappearance signal \cite{tension}. Limited experimental data are available on $\nu_{\mu}$ disappearance at SBL: the dated CDHS experiment \cite{CDHS} and the more recent results from MiniBooNE \cite{MiniBooNE}, a joint MiniBooNE/SciBooNE analysis \cite{MiniBooNE-SciBooNE} and MINOS and SK \cite{MINOS-SK}. The latter results only slightly extend the disappearance exclusion region set by CDHS and currently the region sin$^{2}2\theta_{new}<$ 0.1 is largely still uncostrained.
\\After the decision of CERN not to pursue for the time being the construction of a new neutrino beam, the NESSiE Collaboration proposed to perform an accurate measurement of $\nu_{\mu}$ disappearance at small L/E using magnetic spectrometers (B = 1.5 T) at NEAR and FAR sites on the FNAL-Booster neutrino beam \cite{proposal, papersubmitted}. Many detector configurations were considered investigating experimental aspects such as the measurements of the lepton charge on event-by-event basis and its energy over a wide range. The basic NESSiE concept is to design, construct and install two spectrometers at two sites, ``Near'' (at 110 m, on-axis) and ``Far'' (at 710 m, on surface), in line with the FNAL-Booster, fully compatible with the proposed LAr detectors. Using very massive detectors (300 tons and 700 tons for the Near and Far spectrometers, respectively) it would be possible to collect very large statistical samples in order to improve the disentangling of systematic effects and to span the oscillation mixing parameter down to till un-explored regions (sin$^{2}2\theta_{new}>$ 0.01 or less). 
\\By keeping the systematic error at the level of 1-2\% for the measurements of the $\nu_{\mu}$ interactions, it would be possible to:
\begin{itemize}
\item measure $\nu_{\mu}$ disappearance in the almost entire available momentum range (p$_{\mu}\geq$ 500 MeV/c).
\item measure the neutrino flux at the Near detector, in the relevant muon momentum range, to keep the systematic errors at the lowest possible values;
\item measure the sign of the muon charge to separate $\nu_{\mu}$ from $\bar{\nu}_{\mu}$ for systematics control.
\end{itemize}

The uncertainty on the absolute flux at MiniBooNE stays below 20\% for energies below 1.5 GeV while it increases drastically above that energy. The uncertainty is dominated by the knowledge of hadronic interactions of protons on the Be target, which affects the angular and momentum spectra of neutrino parents emerging from the target. The result is based on experimental data obtained by the HARP and E910 collaborations \cite{HARP-E910}.
\\Such a large uncertainty makes the use of two or more identical detectors at different baselines mandatory when searching for small disappearance phenomena. The ratio of the event rates at the Far and Near detectors (FNR) as function of neutrino energy is a convenient variable since it benefits at first order from cancellation of common proton-target and neutrino cross-sections systematics and of the effects of reconstruction efficiencies. Thanks to these cancellations the uncertainty on the FNR or, equivalently, on the Far spectrum extrapolated from the Near spectrum is usually at the percent level ranging in the 0.5-5\% interval. In order to understand how the hadroproduction uncertainty affects the accuracy on FNR for the specific case of our proposal we developed a new beamline simulation using  either FLUKA, GEANT4 or the Sanford-Wang parametrization for the simulation of p-Be interactions \cite{HARP-E910}.

\section{Physics analysis and sensitivity}
\begin{figure}[!ht]
\begin{center}
\includegraphics[width=0.6\columnwidth]{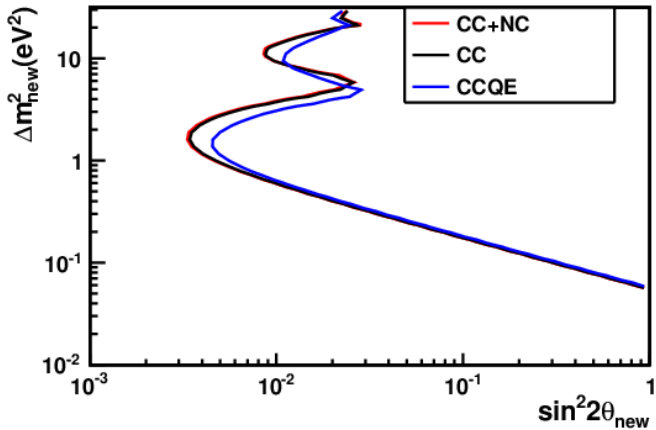}
\caption{95\% C.L. sensitivity obtained using the muon range for all CC (black) and CC+NC (red) events and for only CCQE events (blue) assuming 3 years of data taking. 1\% bin-to-bin correlated error in the shape is considered.}
\label{fig:figure6_sensitivity}
\end{center}
\end{figure}

We developed sophisticated analyses to determine the sensitivity region that can be explored with an exposure of 6.6$\times$10$^{20}$ p.o.t., corresponding to 3 years of data taking on the FNAL-Booster beam. Our guidelines were the maximal extension at small values of the mixing angle parameter, and its dependence on systematic effects. To this aim the sensitivity of the experiment was evaluated using three different analyses implementing different techniques and approximations:
\begin{itemize}
\item a Feldman\&Cousins technique with ad hoc systematic errors added to the muon momentum distribution;
\item a full correlation matrix based on the full Monte Carlo simulation including the reconstruction of the simulated data;
\item CLs profile likelihood;
\end{itemize}
For all analyses the two flavor neutrino mixing in the approximation of one mass dominance is considered. The oscillation probability is given by the formula:
\begin{equation}
P = sin^{2}(2\theta_{new})sin^{2}(1.27\Delta m^{2}_{new}L(km)/E(GeV))
\end{equation}
where $\Delta m^{2}_{new}$ is the mass splitting between a new heavy-neutrino mass-state and the heaviest among the three SM neutrinos, and $\theta_{new}$ is the corresponding mixing angle.
\\In the second method we implemented different smearing matrices for two different observables evaluated with GLoBES \cite{globes}, the muon range and the number of crossed planes at Near and Far detectors respectively, associated with the true incoming neutrino energy. These matrices were obtained through a full Monte Carlo simulation.
In this analysis the correlations between the data collected in the Far and Near detectors are taken into account through the covariant matrix of the observable. 
\\The $\nu_{\mu}$ disappearance can be observed either by a deficit of events (normalization) or, alternatively, by a distortion of the observable spectrum (shape) which are affected by systematic uncertainties expressed by the normalization errors matrix and the shape errors matrix, respectively. The normalization errors matrix is the component of error matrix which is the same for each element. The shape errors matrix represents a migration of events across the bins. 
\\By applying the frequentist method the $\chi^{2}$ statistic distribution was used to determine the sensitivity to oscillation parameters. Sensitivity plots were computed by introducing bin-to-bin correlated systematic uncertainties as expressed in the covariance matrix by considering either 1\% correlated error in the normalization or 1\% correlated error in the spectrum shape. As a representative result the sensitivity calculated considering 1\% correlated error in the shape is plotted in Fig. \ref{fig:figure6_sensitivity}.  The sensitivity computed considering CC and NC events is almost the same as the sensitivity obtained with only CC events \cite{Proposal} and therefore NC background events do not affect the result.

\section{Summary}

Existing anomalies in the neutrino sector may hint to the existence of one or more additional sterile neutrino families. The NESSiE collaboration performed a detailed study of the physics case in order to set a Short-Baseline experiment at the FNAL-Booster neutrino beam to exploit the measurement of the charged current events.
\\The best option in terms of physics reach and funding constraints is provided by two spectrometers based on dipoles iron magnets, at the Near and Far sites, located at 110 and 710 m from the FNAL Booster neutrino source, respectively.
\\ By the proposed approach systematic errors can be kept at the level of 1-2\% for the measurements of the $\nu_{\mu}$ interactions. In particular the measurement of the muon momentum can be performed at the percent level and the identification of its charge on event-by-event basis can be extended to well below 1 GeV.


\bigskip

\begin{thebibliography}{99}


\bibitem{sterile}
B. Pontecorvo, Zh. Eksp. Teor. Fiz. 53, 1717 (1967) [Sov. Phys. JETP 26, 984 (1968)].
\bibitem{tension}
 J. Kopp, P. A. N. Machado, M. Maltoni, and T. Schwetz, JHEP 05, 050 (2013), arXiv:1303.3011;
\\T. Schwetz, Nuclear Physics B, vol. 235-236, pp. 2292235, 2013;
\\C. Giunti, M. Laveder, Y. F. Li, Q. Y. Liu, and H. W. Long, Phys. Rev. D, vol. 86, no. 11, Article ID 113014, 2012.
\bibitem{CDHS}
CDHS Collaboration, F. Dydak \textit{et al.}, Phys. Lett. {\bf B134}, 281 (1984).
\bibitem{MiniBooNE}
MiniBooNE Collaboration, A. A. Aguilar-Arevalo \textit{et al.}, Phys. Rev. Lett. {\bf 103}, 061802 (2009), arXiv:0903.2465.
\bibitem{MiniBooNE-SciBooNE}
MiniBooNE and SciBooNE Collaborations, K. B. M. Mahn \textit{et al.}, Phys. Rev. D85, 032007 (2012), arXiv:1106.5685;
\\G. Cheng \textit{et al.}, Phys. Rev. {\bf D86}, 052009 (2012), arXiv:1208.0322.
\bibitem{MINOS-SK}
 A. Sousa (MINOS) and R. Wendell (Super-K) contributions at Neutrino2014, Boston, USA, 2-7 June 2014.
\bibitem{proposal}  A. Anokhina et al. (NESSiE Collaborations), FNAL-P-1057 (2014), arXiv:1404.2521.
\bibitem{papersubmitted} A. Anokhina et al. (NESSiE Collaboration), arXiv: 1503.07471 [hep-ex].
\bibitem{HARP-E910}
 J. R. Sanford and C. L. Wang (1967), BNL Internal Report, BNL-11479;
\\MiniBooNE Collaboration, A. A. Aguilar-Arevalo et al., Phys. Rev. {\bf D79}, 072002 (2009), arXiv:hep-ex/0601022v1 (2006).
\bibitem{globes}
http://www.mpi-hd.mpg.de/personalhomes/globes/index.html
\bibitem{Proposal}
NESSiE Collaborations, A. Anokhina \textit{et al.}, FNAL-P-1057, arXiv:1404.2521.


\end{thebibliography}
\end{document}